\documentclass[aps,prl,twocolumn,showpacs]{revtex4}

\newcommand{\ie}{i.e. \ }
\newcommand{\eg}{e.g. \ }

\usepackage{graphicx}

\begin{document}

\title{Dynamical instability in kicked Bose-Einstein condensates}

\author{J.~Reslen,$^1$ C.E.~Creffield,$^{1,2}$ and T.S.~Monteiro$^1$}
\affiliation{$^1$Department of Physics and Astronomy,
University College London, Gower Street, London WC1E 6BT, United Kingdom \\
$^2$Departamento de F\'isica de Materiales,
Universidad Complutense de Madrid, E-28040, Madrid, Spain}

\date{\today}

\begin{abstract}
Bose-Einstein condensates subject to short pulses (`kicks')
from standing waves of light represent a nonlinear analogue of the 
well-known chaos paradigm, the quantum kicked rotor. 
Previous studies of the onset of dynamical instability
(ie exponential  proliferation of non-condensate particles)
suggested that the transition to instability might be associated with
a transition to chaos. Here we conclude instead
that instability is due  to resonant driving of Bogoliubov
modes. 
We investigate the Bogoliubov spectrum for
both the quantum kicked rotor (QKR) and a variant, the 
double kicked rotor (QKR-2). We present an analytical model,
valid in the limit of weak impulses which correctly gives the
scaling properties of the resonances and yields good
agreement with mean-field numerics. 
\end{abstract}


\maketitle

\section{Introduction}

The production of Bose-Einstein condensates (BECs) in dilute
atomic gases has opened up a new domain for research in quantum dynamics,
since BECs are intrinsically phase-coherent and can be controlled 
experimentally to an extremely high degree of precision \cite{Stringari}.
An increasingly interesting aspect of the dynamics of BECs is that
they represent a new arena for investigation
of the interaction between nonlinearity and quantum dynamics, including quantum chaos
\cite{Shep,Gardiner,Garreau,Duffy,Zhang,Zhang2,Wimberger,Adams}.

A BEC subject to periodic short pulses, or kicks, from standing waves of
light represents a nonlinear generalization of the 
well-known chaos paradigm, the quantum kicked rotor (QKR).
The QKR has been realized using (non-condensed) cold atoms,
permitting experimental investigation of a range of interesting
chaos phenomena \cite{Raizen}. The regime where the kick-period $T$
is a rational multiple of $\pi$ has also proved of particular
interest: several studies
have investigated the dynamics here with or without
linearity \cite{Darcy,Fish,Wimberger,Rebuzzini}. A number of experimental studies
have also investigated kicked BECs \cite{Sadgrove}.   
Ensuring dynamical stability of the condensate is thus very
important in studies of its coherent dynamics:
if the condensate is dynamically unstable, numbers of
non-condensate particles grow exponentially. If it is stable, they grow
more slowly (polynomially). More broadly,
the study of diffferent types of instability in static \cite{Wu} and driven BECs 
\cite{Dalfovo} is of much current interest.
 
 Previous work on kicked systems\cite{Gardiner,Zhang,Zhang2} considered the onset of
dynamical instability and investigated the relation with
classical chaos. In \cite{Gardiner}, the possibility that instability is
related to chaos in the one-body limit was investigated
for the Kicked Harmonic Oscillator.
In \cite{Zhang,Zhang2} the correlation between
chaos in the mean-field dynamics, rather, and
  the onset of dynamical instability, was investigated.
An  ``instability border'', determined by the
kick strength $K$ and the nonlinearity $g$ was mapped out; 
it was then found \cite{Zhang2} that the parameter ranges for this border
corresponds closely to a transition from regular to chaotic motion,
of an effective classical Hamiltonian derived from the mean-field
dynamics. Hence, present understanding of onset of dynamical instability
in kicked BECs suggests that it may somehow be related to a transition to chaos. 

In this work, we conclude that a quite different mechanism is primarily responsible
for dynamical instability in the QKR-BEC. 
Our key finding is that it is  the strong resonant driving of
certain condensate modes by the kicking, which triggers loss of stability 
of the condensate. This mechanism is unrelated to the transition to chaos,
but is rather an example of parametric resonance. In another context,
the relationship between parametric
resonance and dynamical instability of a BEC in a time-modulated trap is a topic
of much current theoretical \cite{PRTHEO,Dalfovo}  and experimental interest \cite{PREXPT}.
But to date, ``Bogoliubov spectroscopy'' in the analogous time-dependent system, the 
$\delta$-kicked BEC,  has not been investigated.  Our study 
 shows that the temporally kicked BECs open up many new possibilities in this arena.

We find that in general, for the kicked-BEC,
 there is no single stability border: typically, for moderate $K$,
the condensate restabilizes just above the stability border. 
For small $K$ and $g$ the number of non-condensed 
atoms $N_{ex}(t)$ grows exponentially only very close to a few, isolated
resonance peaks.
 With increasing $K$ and $g$, the number of resonances which 
can be strongly excited by the kicking proliferates  and overlaps.
Our calculations show this is associated with generalized
 exponential instability; however this regime is, to a
large degree, beyond the scope of our methods. 
For lower $K$ and $g$, though, we introduce a simple perturbative model which 
provides the approximate position and
width of the important resonances for both rational and irrational $T$.

A key finding is that, for the integer values of $T/\pi= m$ 
(where $m$ is integer) values, the focus of the study in \cite{Zhang},
the onset of instability can occur at nonlinearities  much lower than those
required to resonantly excite even the very lowest collective mode -- a key reason 
why the mechanism of parametric resonance may
so far been overlooked in respect of destabilization of kicked BECs.
Our model demonstrates that for this case, resonant excitation involves
{\em two} excited modes in addition to the initial ground state mode.
Hence we can explain the position of the critical stability border
found in \cite{Zhang,Poletti}.

We investigate both the usual QKR-BEC 
as well as a simple modification, obtained by applying a series of pairs 
of closely-spaced opposing kicks (the QKR2-BEC). This modifies substantially
the relative strengths of the resonances, and provides the added novelty
that the lowest modes are excited by an effective imaginary kick-strength.
It is closely related to the double-kicked quantum rotor, investigated
in cold atoms experiments and theory \cite{Jones}.
We introduce a simple analytical model based on the properties of the 
unperturbed condensate, which gives the distinctive properties and scaling 
behavior of the condensate oscillations on and off resonance.

In Section II we introduce briefly the kicked and double-kicked BEC
systems.
In Section III we introduce the time-dependent Bogoliubov method 
proposed by Castin and Dum and present numerics for
the growth of non-condensate atoms. In Section IV we introduce a
simple perturbative model, based on the one period time evolution operator
for a kicked BEC.  In Section V we show that the simple model and
the time-dependent Bogoliubov numerics give excellent agreement
in the limit of weak kicks. In Section VI we consider the case 
$T=2\pi$ with both numerics and the perturbative model 
and show that the instability border found in \cite{Zhang,Zhang2} is
due to a novel type of compound Bogoliubov resonance.

\section{Kicked BEC systems}

As in \cite{Zhang}, we consider a BEC confined in a ring-shaped trap of 
radius $R$. We assume that the lateral dimension $r$
of the trap is much smaller than its circumference, and thus we
are dealing with an effectively 1D system \cite{rescalg}. The dynamics of
the condensate wavefunction at temperatures well below the transition
temperature are then governed by the 1D Gross-Pitaevskii (GP)
Hamiltonian with an additional kicking potential:
\begin{equation}
H= H_{GP} + K \cos \theta \ f(t) ,
\label{gpe}
\end{equation}
where
\begin{equation}
 H_{GP}=- \frac{\hbar^2}{2 m R^2} \frac{\partial^2}{\partial \theta^2} 
+ g | \psi(\theta,t) |^2 .
\end{equation}

The short-range interactions between the atoms in the condensate
are described by a mean-field term with strength $g=8 N_{tot} a_S R/r^2$,
where $a_S$ is the s-wave scattering length, and $N_{tot}$ is
the total number of atoms.
For the QKR-BEC system,  $f(t)= \sum_n  \delta(t - nT)$, while for the 
QKR2-BEC, 
\begin{equation}
f(t) = \sum_n \left[ \delta(t - nT) \ - \ \delta(t- nT+\epsilon ) \right] ,
\label{kick}
\end{equation}
where $T$ is the total period of the driving; $\epsilon \ll T$ 
and thus the second kick nearly cancels the first.\\
 
Experimental and theoretical studies of the double-kicked rotor 
\cite{Jones} have shown that its quantum behavior is 
largely determined by an effective kick strength $K_\epsilon= K\epsilon$,
provided $T \gg \epsilon$.
Here we take $\epsilon=1/25$. Hence, while for the QKR-BEC, the value
$K=1$ represents a relatively large impulse for a kicked BEC,
for a double kicked BEC, $K=1$ in the numerics below corresponds to 
$K_\epsilon=0.04$, and represents only a very weak impulse. 
The reason for this is the near cancellation of consecutive
kicks in each pair.

This mechanism has certain analogies with the so-called ``quantum antiresonance''
investigated in \cite{Zhang}: for QKRs kicked at $T =2\pi$, 
consecutive kicks effectively cancel. This means that 
even large values of $K \simeq 1$ and $g > 1$ represent
only weak driving; for example,  the instability border was found by
\cite{Zhang} to occur at $g \simeq 2$ and $K=0.8$.

\section{time-dependent Bogoliubov method}
The number of non-condensed atoms were calculated by making the 
usual Bogoliubov approximation, and following the formalism of 
Castin and Dum \cite{castin}. This adaptation of the Bogoliubov 
linearization for time-dependent potentials has been used in
all studies to date of the dynamical stability of kicked
condensates \cite{Gardiner,Zhang,Poletti,Rebuzzini}.
The mean number of non-condensed atoms at zero temperature is given by
$N_{ex}(t) = \sum_{k=1}^{\infty} \langle v_k(t) | v_k(t) \rangle$,
where the amplitudes $(u_k, v_k)$ of the Bogoliubov
quasiparticle operators are governed by the coupled equations 
\begin{equation}
i \hbar \frac{d}{dt}\left(\! \!
\begin{array}{c}
u_k\\
v_k
\end{array}
\! \! \right) =\left(\! \!
\begin{array}{cc}
H+gQ|\psi |^2Q &\! gQ\psi ^2Q^* \\
-gQ^*\psi^{*2}Q &\! -H-gQ^*|\psi |^2Q^*
\end{array}
\! \!\right)
\!\left(\! \!
\begin{array}{c}
u_k  \\
v_k
\end{array}
\! \!\right) .
\label{bogo}
\end{equation}
In this expression, $Q = I - | \psi \rangle \langle \psi |$
are projection operators that orthogonalize the quasiparticle modes 
with respect to the condensate \cite{castin}. 
We assume that at time $t=0$, we have a homogeneous condensate
$\psi_0 =1/\sqrt{2\pi}$. Further discussion of the theory is given in
\cite{gard}.

The regime of validity of the method is discussed in \cite{castin}.
 The method is valid in the weakly interacting limit $1 \gg a_s^3 \rho$
where $\rho$ is the density. A limit is identified where this condition
is satisfied, if one works with a constant $g \propto  N_{tot} a_s$; thus the limit
$a_s \to 0$ corresponds to $N_{tot} \to \infty$.  
A further requirement is that condensate depletion remains negligible.
This condition fails after a few kicks in exponentially unstable regions.
Here the method is employed only to identify the  the parameter range for
the onset of instability.
We cut-off our calculations for $N_{ex} > 10^3$ (a reasonable threshold for
small depletion in a condensate with $N_{tot} \sim 10^5$).

\begin{figure}[htb]
\includegraphics[width=2.5in]{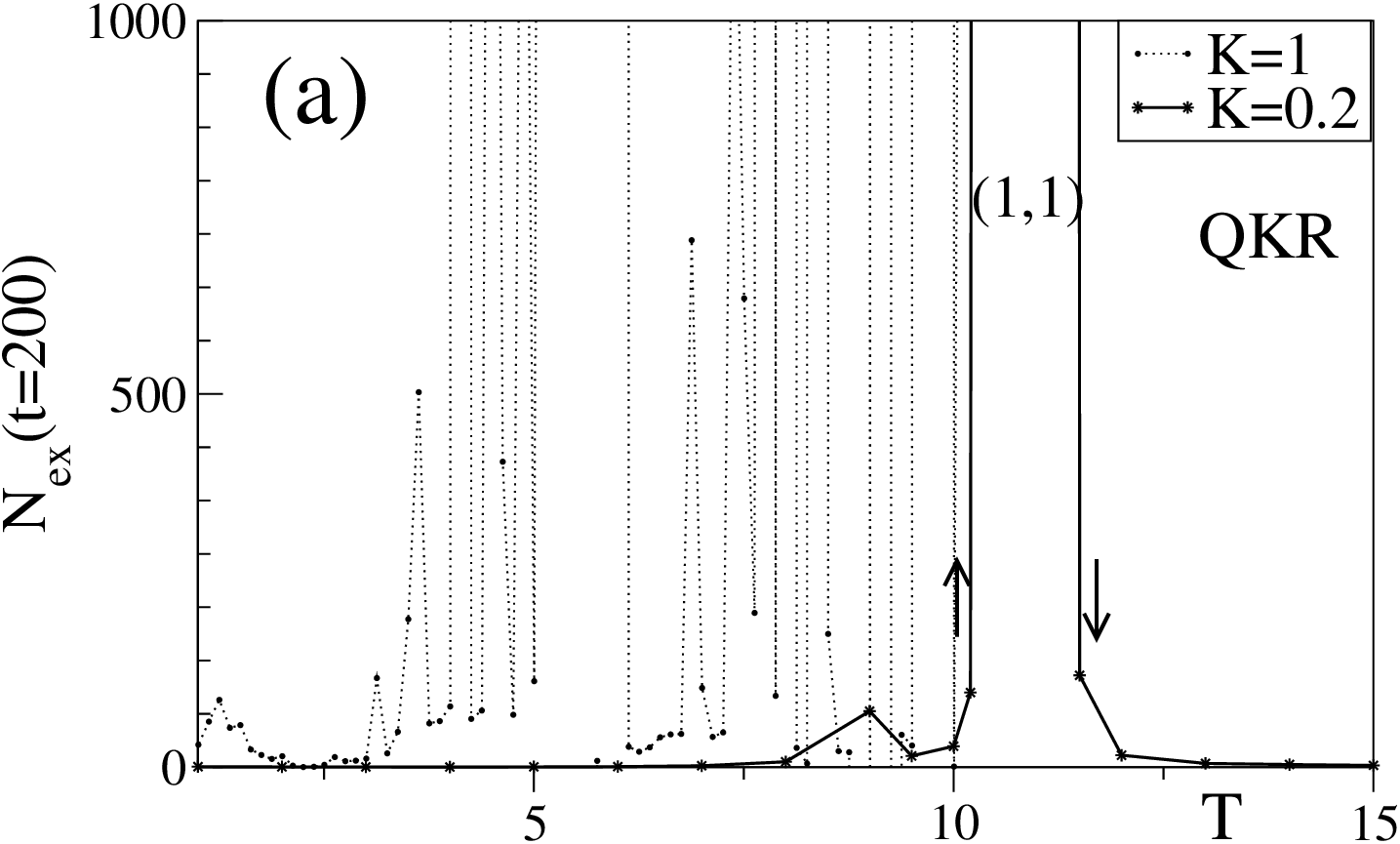}
\vspace*{10 mm}
\includegraphics[width=2.5in]{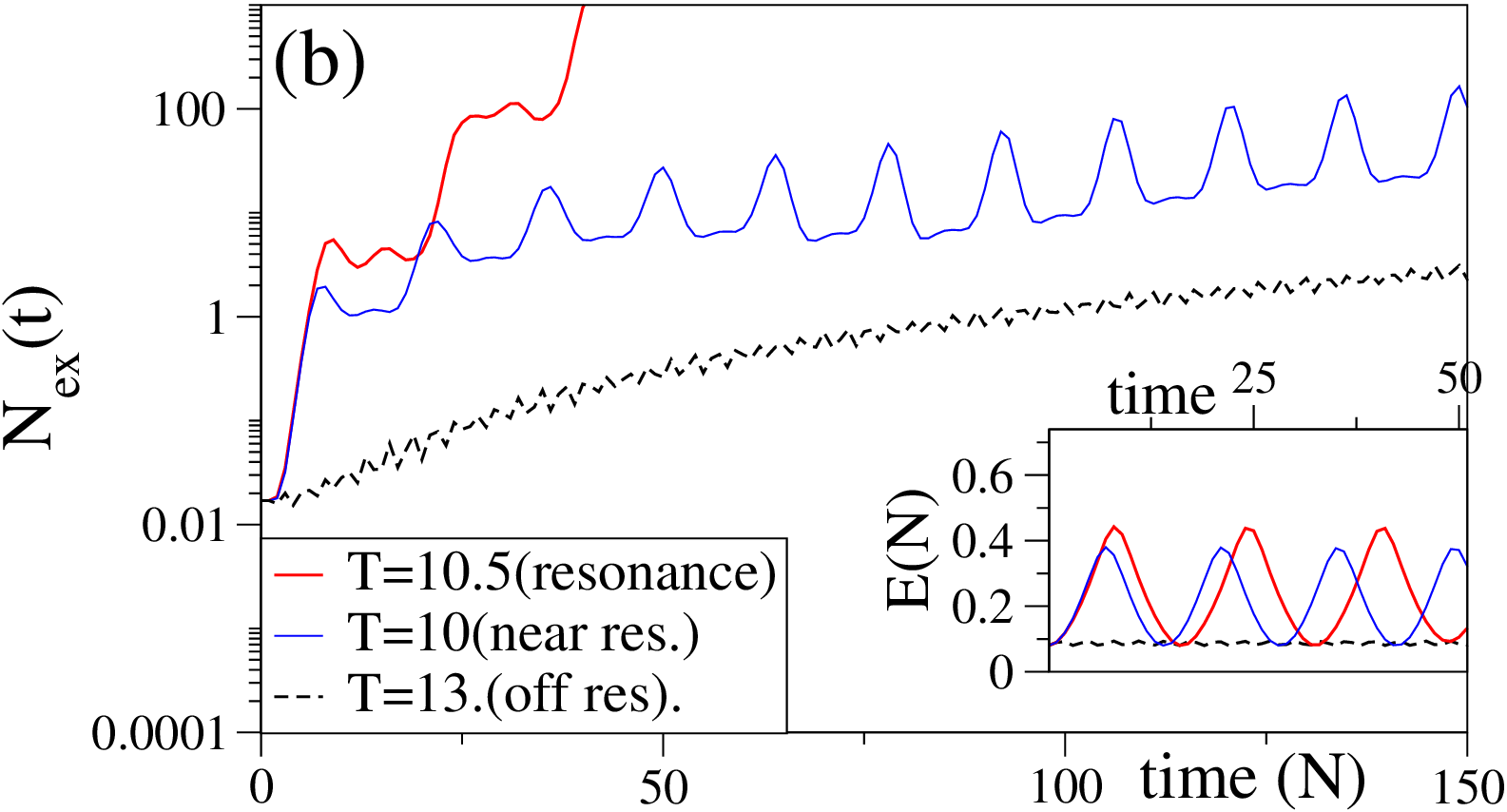}
\vspace*{-10 mm}
\caption{{\bf(a)} Shows that for weak kicks (solid line), instability occurs only at one isolated
 Bogoliubov $(1,1)$ resonance , where $(n,l)$ denotes the $n-th$ resonance of eigenmode $l$.
 ``Up'' arrows indicate
onset of exponential instability; `` down'' arrows means stability is regained.
 $g=1$. The total number of 
non-condensate atoms generated after 200 kicks, $N_{ex}(N=200)$,
is plotted as a function of kicking period $T$.
 For stronger kicks (dotted line; $K=1$,$T<10$) resonances
proliferate and there is instability over almost all the parameter range. 
{\bf (b)} Time-dependence near the$(1,1)$ resonance at $T \approx 10$
corresponding to Fig (a).  
Non-resonant ($T=13$) curve shows weak quasi-periodic oscillations
in $N_{ex}$; the near-resonant regime, $T=10$ 
is characterized by slow, large oscillations; at resonance $T = 10.5$,
there is exponential growth in $N_{ex}(t)$.
Inset shows that the condensate energy (calculated from the GPE itself) has similar
oscillations.}
\label{Fig1}
\end{figure}

In Figs.\ref{Fig1} (a) and (b) we show the number of non-condensed atoms,
$N_{ex}(t=NT)$, calculated from the Bogoliubov
equations (\ref{bogo}) after $N=200$.
For small $K=0.2$, $g=1$,  a single resonance is seen
at $T \simeq 10$. 
For small $K$, resonances  occur whenever the resonance 
condition \cite{Dalfovo} 
$\omega_0+\omega_{{l}} = \omega_{{l}} \approx \frac{2n\pi}{T} $ is satisfied,
 where $n=1,2,3..$ is an integer and $\omega_{{l}}$ is the eigenfrequency
of the $l-th$ collective mode. 
For larger $K=1$, the figure shows that 
resonances are extremely dense and overlap with each
other (and we show the behavior in this regime
for $T<10$). For overlapping resonances, unambiguous
identification of each resonance is no longer possible.
The key point here, however, is that in the stable regions outside the resonances, 
$N_{ex}$ remains very small even after prolonged kicking.\\

Fig.\ref{Fig1}(b) 
shows   oscillations of $N_{ex}$, as a function of time, for weak $K=0.2$, $g=1$,
close to the isolated resonance at $T \approx 10$. The three possible 
regimes of: (non-resonant) weak quasi-periodic oscillations in time;
(near-resonant) slower, large periodic  oscillations; 
and (resonant) exponential growth are illustrated. 
The condensate energy,  
$E(N)=\int_0^{2\pi} d\theta \psi^*(N) (-\frac{1}{2} \frac{\partial^2}{\partial \theta^2} +\frac{g}{2}|\psi(N)|^2) \psi(N)$
after $N$ kicks, obtained from the GPE itself, is also shown, for 
comparison, in the inset: at resonance,  large oscillations are also seen.\\

\begin{figure}[htb]
\includegraphics[width=3in]{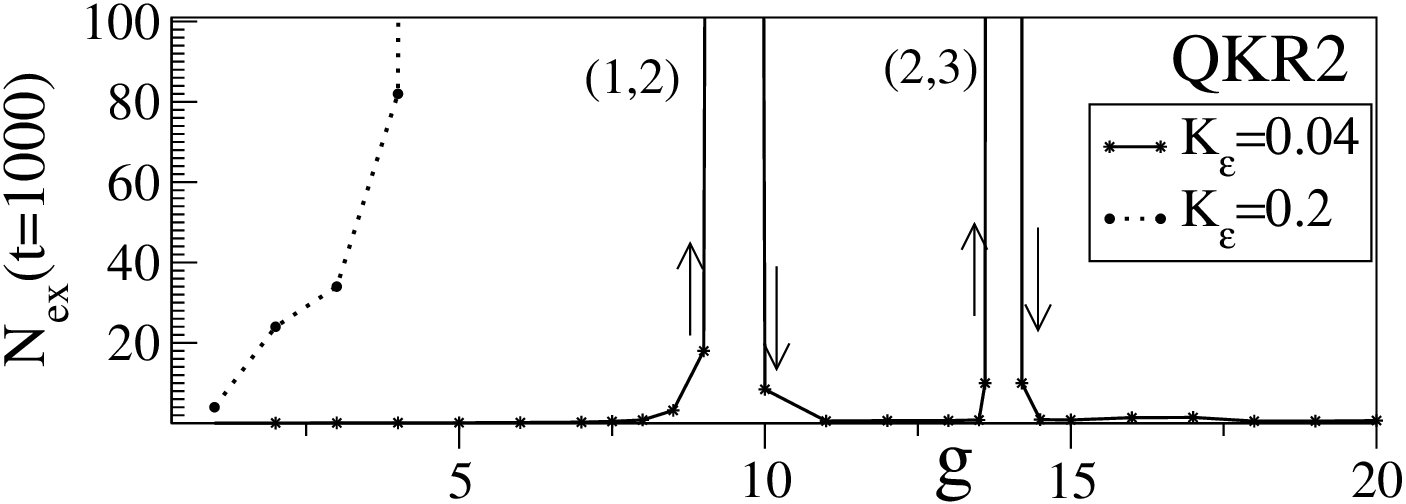}
\caption{double-kicked BEC (QKR2): Shows zones of instability occur  
at Bogoliubov resonances. Condensate losses 
as a function of nonlinearity parameter $g$.
 ``Up'' arrows indicate
onset of exponential instability; `` down'' arrows means stability is regained.
 $N_{ex}(t=1000)$ is plotted as a function of $g$ (for $T=2$, $\epsilon=1/25$) for
weak kicks ($K=1$ so effective kick is $K_\epsilon=0.04$) and stronger
kicks ($K=5$ so effective kick $K_\epsilon=0.2$).}
\label{Fig2}
\end{figure}

Fig.\ref{Fig2} shows the corresponding behavior for the double-kicked
BEC, but now as a function of $g$, keeping $T=2$, $\epsilon=1/25$ constant
and $K=1$ or $K=5$ (hence $K_\epsilon=0.04$ or $0.2$).
The curve $K_\epsilon=0.04$ corresponds to weak impulses and shows two isolated
Bogoliubov resonances. While values of $g \simeq 10$ are large compared
with current experimental values of $g \sim 0.5$ (see discussion of experimental
$g$ in \cite{Rebuzzini}), resonances at small $g \sim 1$ more suitable for 
experimental spectroscopy can be excited by considering larger $T$.
 The curve $K_\epsilon=0.2$ is in the overlapping
resonance regime, so produces generalised instability.

In order to understand the behavior at the resonances, we 
introduce in Section II  a  model for the time evolution of perturbations
from the kicked condensate, based on the usual linearization with respect to
small perturbations.

\section{II: Kicked condensate model}

The time-evolution of small perturbations of the condensate itself 
are described by an equation similar to Eq.\ref{bogo}, see  \cite{castin}.
We write the condensate wavefunction
in the form $ \psi=\psi_0 +\delta \psi$, where  $\psi_0$ is the 
unperturbed condensate and $\delta \psi$ represent 
the excited components. Inserting this form in the GPE
and linearizing with respect to $\delta \psi$,
we can write:
\begin{equation}
i \hbar \frac{d}{dt}\left(\! \!
\begin{array}{c}
\delta \psi\\
\delta \psi^*
\end{array}
\! \! \right) = {\cal L}(t)
\!\left(\! \!
\begin{array}{c}
\delta \psi\\
\delta \psi^*
\end{array}
\! \!\right) .
\label{bec}
\end{equation}

where,
\begin{equation}
{\cal L}(t) =
\left(\! \!
\begin{array}{cc}
H(t) +g|\psi |^2 &\! g\psi^{*2} \\
-g\psi^{*2} &\! -H(t) -g|\psi |^2
\end{array}
\! \!\right) .
\label{bec}
\end{equation}

The analysis of condensate stability for a time-periodic system
 \cite{Dalfovo} reduces to the analysis of the operator ${\cal L}(t)$
over one period $T$. In general, for systems like BECs in modulated optical lattices,
inter-mode coupling requires a detailed analysis of the instantaneous
evolution. The nature of the
$\delta$-kicked potential permits considerable simplification.

The effect of ${\cal L}(t)$ reduces to the 
 free-ringing of the eigenmodes of the unperturbed condensate
for period $T$, interspersed by instantaneous impulses which mix the modes.
Even for an experiment (where the kicks are approximated pulses of very short, but
finite duration) numerical time-propagation is avoided: intermode
coupling occurs over a very short time-scale, during which eigenmode phases
 remain essentially constant.

Excluding the kick term for the moment, we recall that the time propagation 
under $H_{GP}$ can be analyzed in terms of the eigenmodes $(u_k(t),v_k(t))$ 
and eigenvalues of $\omega_{ k}(t)$ of the
$2 \times 2$ matrix on the right hand side of Eq.\ref{bec}.  
Setting $\psi = 1/\sqrt{2\pi}$,
the matrix can be diagonalized and there are well-known analytical 
expressions for the unperturbed eigenmodes \cite{Stringari}
\begin{eqnarray}
(u_k(t=0),v_k(t=0))= \!\left(\! \!\begin{array}{c}
U_k  \\
V_k
\end{array}
\! \!\right) \frac{e^{ik\theta}}{\sqrt{2\pi}} ,
\label{mode}
\end{eqnarray}
where $U_k+V_k =A_k, \ U_k-V_k =A_k^{-1}$, and
$A_k= \left( \frac{\hbar^2 k^2}{2}(\frac{\hbar^2 k^2}{2}
+ \frac{g}{\pi})\right)^{1/4} $.

In order to understand the behavior at the resonances, we introduce
below a simple model using the eigenmodes Eq.\ref{mode} as a basis.
Writing the small perturbation  in this basis:
\begin{equation}
\!\left(\! \!
\begin{array}{c}
\delta \psi(t)\\
\delta \psi^*(t)
\end{array}
\! \! \right) = \sum_k \ b_k(t)
\!\left(\! \! 
\begin{array}{c}
U_k  \\
V_k
\end{array}
\! \!\right) \frac{e^{ik\theta}}{\sqrt{2\pi}} + b_k^*(t)
\!\left(\! \! 
\begin{array}{c}
V_k  \\
U_k
\end{array}
\! \!\right) \frac{e^{-ik\theta}}{\sqrt{2\pi}} .
\label{basis}
\end{equation}

Neglecting the kick, evolving the modes from some initial time $t_0$,  
each eigenmode $(u_{k},v_{k})$ simply acquires 
a phase ie:
\begin{eqnarray}
b_k(t) = b_k(t_0) e^{-i  \omega_{k} (t-t_0)} ,
\label{umode}
\end{eqnarray}
where $\omega_{k}= \sqrt{\frac{k^2}{2}(\frac{\hbar^2 k^2}{2}+\frac{g}{\pi}})$.

After a time interval $T$, a kick is applied which couples the
eigenmodes. Its effect is obtained by expressing the perturbation in a momentum basis,
$\psi= \sum_{l} a_{l}(t) |{l}\rangle $ where
$|l\rangle = \frac{e^{il\theta}}{\sqrt{2\pi}}$, and we can restrict ourselves
to the symmetric subspace $a_{l}= a_{-l}$ of the initial condensate (parity is
conserved in our system).
Then, we can see by inspection that
\begin{eqnarray}
a_k(t) = U_k b_k(t) + V_k b_{-k}^*(t).
\label{atob}
\end{eqnarray}
Note that $b_k=b_{-k}$ for this system.
Conversely, the corresponding amplitude $b_k$ in each eigenmode $k$ is 
given trivially from Eq.\ref{basis} using orthonormality
of the momentum states and the relation $\ U_k^2- V_k^2=1$, yielding
\begin{eqnarray}
b_k(t) = U_k a_k(t) - V_k a_k^*(t) .
\label{btoa}
\end{eqnarray}

If the evolving condensate is given in the momentum basis, the effect 
of a kick operator $U_{kick}= e^{\pm \frac{i K}{\hbar}\cos \theta }$ is well-known.
The matrix elements:

\begin{eqnarray} 
U_{nl} =\langle n| U_{kick}| l \rangle 
= J_{n-l} (K/\hbar) i^{\pm (l-n)}
\end{eqnarray}
The $J_{n-l}$ are Bessel functions. 

The  amplitudes $a_{l}(t)$ are given by
\begin{eqnarray}
a_n(t^+)= \sum_l \ i^{\pm (l-n)} J_{n-l} \left( \frac{K}{\hbar} \right) \ 
a_{l}(t^-) ,
\label{ukick}
\end{eqnarray}
where $a_n(t^+) / a_l(t^-)$ denotes the amplitude in state $| n \rangle$ 
just after/before the kick.

We can now define a ``time-evolution''  operator \\
${\cal L'}(T) = {\cal B}^{-1}{\cal L}_{free} (T){ \cal B} \  U_{kick}$,
where ${\cal L}_{free}$ denotes free ringing of
the eigenmodes, ${ \cal B}$ is the transformation from momentum
basis to Bogoliubov basis and  $U_{kick}$ is the action of the kick.
A usual procedure for stability analysis of a driven condensate is to
examine the eigenvalues of ${\cal L'}(T)$ to ascertain  whether
there is one (or more eigenvalues) which have a real, positive component
\cite{Dalfovo}, ie whether they produce
exponential growth in the amplitudes $a_{\pm l}$.

However, to compare with GPE numerics, we  simply evolve
the mode amplitudes in time over a few kicks and examine the overall
condensate response to the kicking (in the limit of very weak kicking).
 Hence we can evolve the  amplitudes $a_l(t=NT)$  of the condensate
perturbation from period $N$ to period $N+1$:
\begin{eqnarray}
{\bf a}((N+1)T)= {\cal L'}(T) \ {\bf a}(N) ,
\label{map}
\end{eqnarray}
using only the simple analytical coefficients in Eq.\ref{ukick}
and Eq.\ref{umode}, 
provided we use the simple transformations in Eqs.\ref{atob}
and Eq.\ref{btoa}  to switch between the Bogoliubov mode basis and the momentum basis.
${\cal L'}(T)$ is non-unitary,
but the method is quantitative in the perturbative limit
provided $\psi \simeq \psi_0$, ie we assume $a_0(N)=a_0(0)=1$. 

We calculate the average energy over the first few $N$  kicks, 
$\langle E(N) \rangle = \frac{1}{N} \sum_{t=1}^{N} E(t)$.
Slow, large amplitude oscillations in $E(t)$ yield a large 
$\langle E(t) \rangle$ and indicate 
a resonance. Fig.{\ref{Fig3a}(a) shows the QKR-BEC behavior, 
for equivalent parameters to Fig.{\ref{Fig1}(a). 
For low $K=0.2$, there is the same single $(1,1)$  resonance at $T \approx 10$
as in Fig.{\ref{Fig1}(a).
For higher $K=1$ the method is far from quantitative:
 the model Eq.\ref{map} is only a valid means of time-evolving
the perturbation over a few kicks for  small 
$K<<1$ since it assumes the perturbed component
is negligible; nevertheless, for $K=1$ it illustrates the regime
of dense, overlapping resonances.

In Fig.{\ref{Fig3b}(b) we compare the perturbative Eq.(\ref{map}) results with full GPE numerics 
for the first 20 kick pairs of the QKR2 in the limit of weak kicks. 
 It shows remarkably good agreement.
Moreover the scaling of the resonances with $K$ is well described.
The QKR2 resonant Bogoliubov spectrum differs appreciably from the
QKR case. 
Fig.{\ref{Fig3b}(b) shows that for QKR2, even for low $K=1, \epsilon=1/25$
 ie $K_\epsilon \approx 0.04$ 
and low $g =1$, both $l=1$ and $l=2$ resonances are strongly excited. 
The QKR2-BEC resonance intensity depends strongly on
$K$: the $l=2$ resonances scale as $K^4$, while the $l=1$ scaling is closer to
$K^2$. In the full GPE numerics, the position of the maxima depends slightly 
on $K$ and $g$, but remains within a few percent of the unperturbed value, 
even for longer kicking times if $K_\epsilon$ remains small.\\ 

In the limit of weak driving, one can obtain explicit expressions 
for the condensate wavefunction as a function of time.
We assume that $a_0 \approx 1/{\sqrt 2 \pi} \gg a_{l\neq 0}$.
Then Eq.\ref{ukick} can be approximated by
$a_n(t^+) \approx  a_{l}(t^-) + U_{l0}/({\sqrt 2 \pi})$. 
From Eq.\ref{btoa} and Eq.\ref{umode} we see that the amplitude 
accumulated over a single period in
each eigenmode is 
\begin{eqnarray}
b_l(N+1)= b_l(N) + (U_l U_{l0} -V_l U^*_{l0})  e^{i \omega_l T} .
\end{eqnarray}
Summing all contributions iteratively from $t=0$, taking $b_l(0)=0$, we obtain
\begin{eqnarray}
b_l(N)=(U_l U_{l0} -V_l U^*_{l0}) \sum_{n=0}^{n=N-1} e^{i n\omega_l T} ,
\end{eqnarray}
and so for $\omega_l T \approx 2\pi$ all the contributions add in phase, 
analogously to the well-known (but unrelated) resonances of the non-interacting
limit \cite{Darcy}.

We can write $\sum_{n=0}^{N-1} e^{i n\omega_l T}=e^ {-i (N-1) \omega_l T} 
\Phi(\frac {N\omega_l T}{2})$ where the $\Phi$ function is: 
\begin{eqnarray}
\Phi \left( \frac {N\omega_l T}{2} \right) = 
\frac{\sin \left( N\omega_l T/2 \right)}{\sin \left( \omega_l T/2 \right)} .
\end{eqnarray}
We thus expect oscillations in each set of $\pm l$ momentum components 
of amplitude
\begin{equation}
|2 a_l(N)|^2  \propto 4 |U_{l0}|^2 \Phi^2(\frac {N\omega_l T}{2}) .
\end{equation}
Off-resonance there will be quasi-periodic oscillations (in \eg
the condensate energy) 
from the superposition of contributions characterized by different 
eigenfrequencies $\omega_l$.
Close to resonance, a single component dominates; if the $l-th$ mode
is resonant we can write $\omega_l T \approx 2\pi M + 2\delta$ 
where $2\delta \ll 1$ is the de-phasing from resonance. 
Then 
\begin{equation}
|a_l(N)|^2  \propto \frac{|U_{l0}|^2}{\delta^2} \sin ^2 (N \delta) ,
\label{res}
\end{equation}
and there are slow, periodic oscillations of large amplitude 
$\sim 4 \frac{|U_{l0}|^2}{\delta^2}$,
at a frequency $\delta$ which is {\em not} related to any eigenmode frequency,
but given rather by the de-phasing from resonance.

The QKR2 resonant excitation spectrum is rather different 
from the QKR, and is analysed further
in the next section.

\section{III: Resonances of the QKR2-BEC}
In the limit $K_\epsilon \to 0$, we can obtain analytical expressions
for the BEC wavefunction of the double-kicked system.
Firstly note that when $g\epsilon \ll 1$, 
the non-linearity has little effect during the
short time-interval $\epsilon$. Using the relation,
\begin{eqnarray}
 e^{+i \frac{K}{\hbar} \cos \theta} \ e^{-ip^2 t\hbar/2} e^{-i \frac{K}{\hbar} \cos (\theta)}= 
e^{-\frac{i t}{2\hbar}\left[\hat{p} + K \sin \theta\right]^2},
\end{eqnarray}
the time evolution can be given as a `one-kick' operator
\begin{eqnarray}
{\hat U}(T) \approx U^{(0)}_{GP}(T,0)
e^{-\frac{i\epsilon}{2\hbar}\left[\hat{p} + K \sin \theta \right]^2} .
\label{onekick}
\end{eqnarray}
In the limit $p \epsilon \approx 0$, one can split the operators in
Eq.\ref{onekick} and neglect a term  $K \sin \theta \ \hat{p}$
to obtain the approximation
\begin{eqnarray}
{\hat U}(T) \approx  e^{-\frac{i}{2\hbar}\hat{p}^2 T} . \ 
e^{-\frac{i}{\hbar}\left[\frac{K^2 \epsilon}{2} \sin^2 \theta 
 - i K \epsilon \hbar \cos \theta \right]} , 
\label{onekick2}
\end{eqnarray}
leaving an effective single-kick quantum rotor with a kicking potential
\begin{eqnarray}
V_{kick} = \left[\frac{K^2 \epsilon}{2} \sin^2 \theta  - 
i K\epsilon \hbar \cos \theta \right]\sum_N \delta (t-NT) .
\label{eq13}
\end{eqnarray}
The second term, curiously,  appears as kicking potential with an 
imaginary, and $\hbar$ dependent, kick strength
$iK \hbar$. It is of purely quantum origin as it arises from 
the non-commutativity of $p$ and $\sin \theta$, \ie
\begin{eqnarray}
i K\hbar \cos \theta =[ K \sin \theta, \hat{p}] .
\label{eq14}
\end{eqnarray}
Nevertheless, as seen below, it is important for weak driving as
it controls the amplitude of the first excited mode $l= \pm 1$.

\begin{figure}
\includegraphics*[width=3.in]{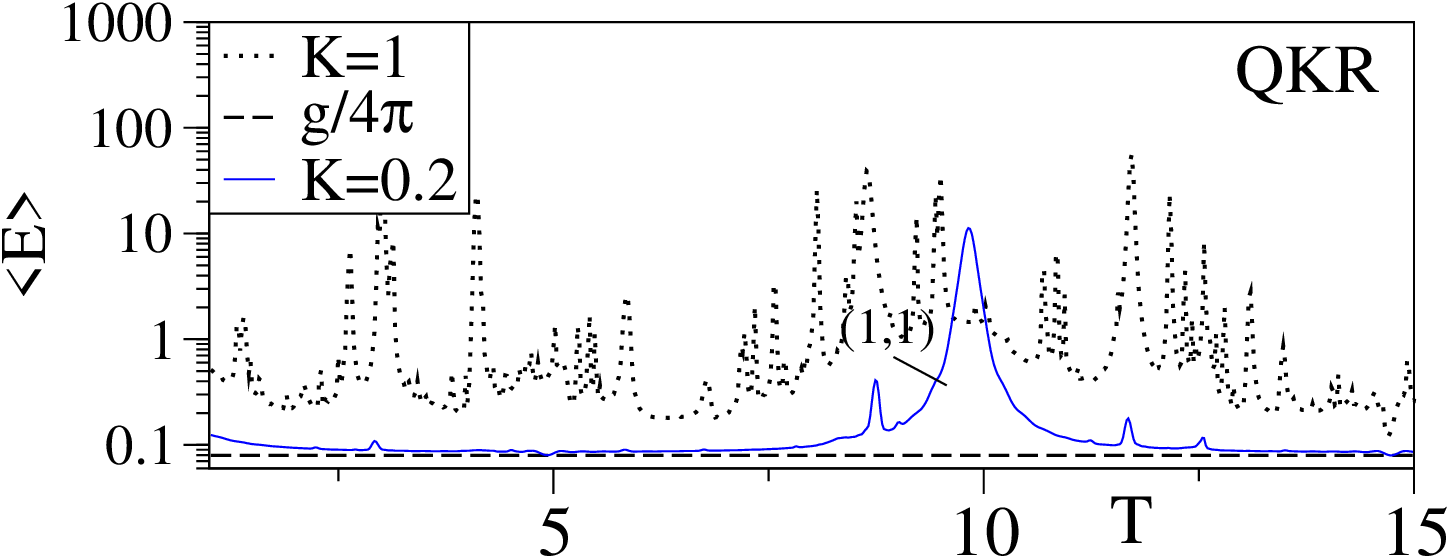}
\caption{Average energy $\langle E \rangle$ after 40 kicks. 
The dashed lines indicate the model of Eq.\ref{map};
all other plots use full numerics.
The label $(n,l)$ denotes $n-th$
resonance of mode $l$.
 Resonances of the QKR-BEC for parameters
comparable to Fig.1a. For low $g=1$, $K=0.2$, only the single 
isolated $(1,1)$ resonance is seen.
For higher  $K=1$, resonances proliferate and overlap.}
\label{Fig3a}
\end{figure}

\begin{figure}
\includegraphics*[width=3.in]{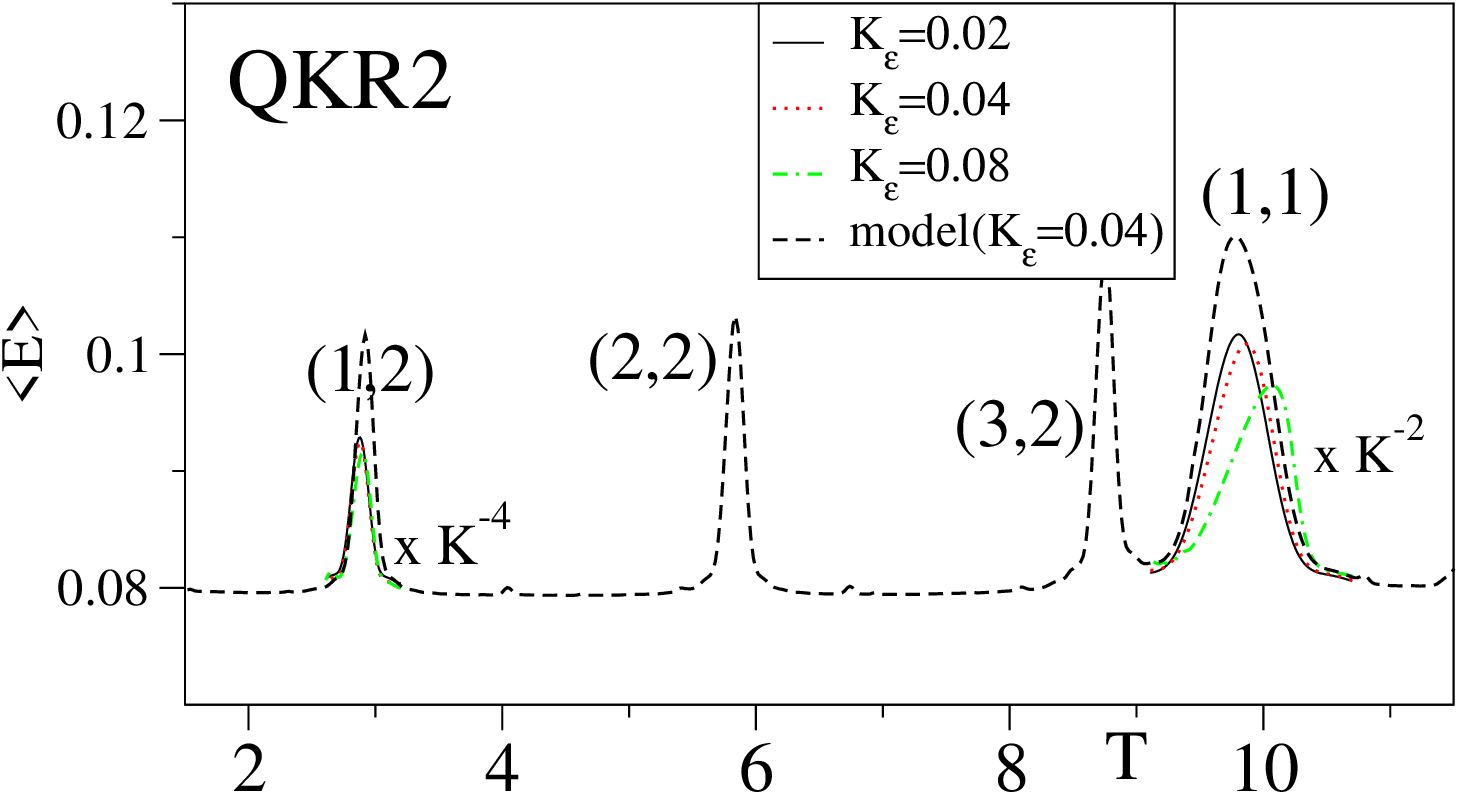}
\caption{Comparison between full GPE numerics and 
the model of Eq.\ref{map} for the QKR2-BEC, showing
excellent agreement. 
Average energy $\langle E \rangle$ after 20 kick-pairs.
The label $(n,l)$ denotes $n-th$
resonance of mode $l$.
$g=1$ and $\epsilon=1/25$ so $K=1$ corresponds to effective
kick strength $K_\epsilon=0.04$. For low $K$, the $l=1$ resonance amplitudes
scale as $\sim K^2$ while those of the $l=2$ modes scale as $\sim K^4$.}
\label{Fig3b}
\end{figure}

The matrix elements of the modified kick $V_{kick}$, like those in  
Eq.\ref{ukick}, are Bessel functions. Specifically, the effect of $V_{kick}$
on the condensate amplitudes $a_l$ is given by 
\begin{eqnarray}
a_n(t^+)= \sum_l U_{nl} \ a_l(t^-) ,
\label{ukick2}
\end{eqnarray}
where $U_{nl}=\sum_m i^{n-l-m} J_m (\frac{K^2 \epsilon}{4 \hbar}) 
J_{n-l-2m} (i K \epsilon)$, and $a_n(t^\pm)$ indicates momentum amplitudes 
before(-) and after(+) the kick, as in Eq.\ref{ukick}.
Since $K \epsilon \ll 1$ and $J_{|n|>1}(z) \simeq 0$, 
only Bessel functions of low order ($m=0$ or $1$) will be non-negligible,
and we can use the small-argument approximations for them, namely
$J_0 (z) \approx 1 $,  $J_{\pm 1}(z) \approx \pm z/2$.

Then, if the condensate is relatively unperturbed, the main effect of the 
kick will be to simply excite a
small amount of $l= \pm 1$ and $l=\pm 2$ from the $|0 \rangle$ state 
\begin{eqnarray}
e^{-\frac{i}{\hbar} V_{kick}} \psi \approx e^{-\frac{i}{\hbar} V_{kick}} \ 
 |0 \rangle = \sum_l U_{l0} |0 \rangle 
\end{eqnarray}
where
\begin{eqnarray}
\sum_l U_{l0} |0 \rangle &\approx& \frac{1}{\sqrt{2\pi}} +
i J_1\left( \frac{i K \epsilon}{2} \right) |\pm 1 \rangle
 + i \ J_1 \left( \frac{K^2 \epsilon}{4 \hbar} \right) |\pm 2\rangle  \nonumber \\
&\approx&
\frac{1}{\sqrt{2\pi}} - \frac{ K \epsilon}{4}|\pm 1\rangle + 
i \ \frac{K^2 \epsilon}{8 \hbar} |\pm 2 \rangle
\end{eqnarray}
We obtain a similar equation to the QKR-BEC for the mode amplitudes, \ie \\ 
$b_l(N)=(U_l U_{l0} -V_l U^*_{l0}) \sum_{n=0}^{N-1} \exp[i n\omega_l T]$.

But if only the lowest excited modes are significant,
then, in particular,\\ $b_1(N)= -\frac{ K \epsilon}{4}(U_1-V_1)
\sum_{n=0}^{N-1} \exp[i n\omega_1 T]$
and \\ $b_2(N)= i \ \frac{K^2 \epsilon}{8 \hbar}(U_2+ V_2)
\sum_{n=0}^{N-1} \exp[i n\omega_2 T]$.
For $\omega_l T \approx 2\pi$
all the contributions add in phase and we will have a resonance of 
either the $l=1$ or $l=2$ modes, the regime illustrated in Fig\ref{Fig2}(b).

Similarly as for the QKR-BEC, we can sum all the contributions to obtain an 
approximate analytical expression for the evolving 
condensate wavefunction including 
excited modes $l=\pm 1$ and  $l=\pm 2$,
\begin{eqnarray}
\psi(N) \approx \frac{1}{2\pi}[1+ C_1 \frac{K\epsilon}{2} \cos \theta +
C_2 \frac{K^2 \epsilon}{4\hbar} \cos 2\theta] .
\label{wave}
\end{eqnarray}
where 
\begin{eqnarray*}
C_1 &=& - \Phi (N {\tilde {\omega_1}})[ \cos{(N-1){\tilde {\omega_1}}}-
i A_1^{-2} \sin{(N-1){\tilde {\omega_1}}}] , \\ 
C_2 &=&  \Phi (N{\tilde {\omega_2}})[A_2^2 \sin{(N-1){\tilde {\omega_2}} }+
i \cos{(N-1){\tilde {\omega_2}}},
\end{eqnarray*}
and ${\tilde {\omega_j}}= \omega_j T/2$.

Eq.\ref{wave} shows that the amplitudes $|a_1|^2$  and $|a_2|^2$ 
scale as $K^2$ and $K^4$ respectively, as seen in the numerics in 
Fig.\ref{Fig3b}(b). Fig.\ref{Fig4}(a) shows that Eq.\ref{wave} gives 
excellent agreement with GPE numerics,
giving accurately the non-resonant quasi-periodic condensate oscillations.
Near the $l=2$ resonance of Fig.\ref{Fig2}, Fig.\ref{Fig4}(b) confirms
the QKR2 condensate oscillations (obtained from the GPE)  scale quite accurately as 
$\propto  \frac{1}{(\delta)^2} \sin^2 {N\delta}$ as expected from 
Eq.\ref{res} and Eq.\ref{wave}.

Fig.\ref{Fig4}(c) shows that, near-resonance, there are 
corresponding large oscillations in the non-condensate numbers calculated
from Eq.\ref{bogo}. 
Near-resonance, $N_{ex}$ increases quadratically with time, on-resonance, 
the increase is exponential.

\begin{figure}
\includegraphics[width=3.in]{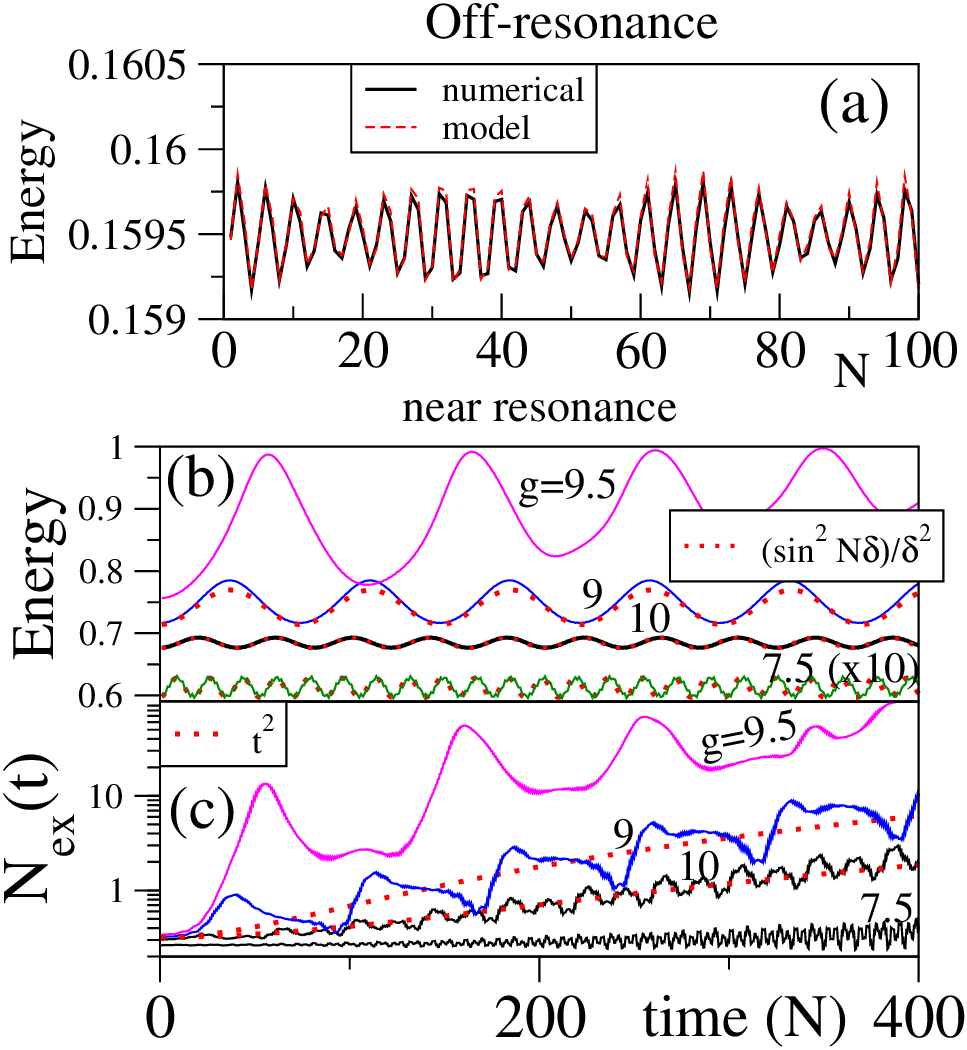}
\caption{Test of perturbative model.
{\bf(a)} Condensate energy oscillations 
from GPE numerics and Eq.\ref{wave}. $K_\epsilon=0.04$, $g=2$, $T=2$. Beating between 
modes 1 and  2 is very accurately described by Eq.\ref{wave}.
{\bf(b)} Behavior of $l=2$ resonance of Fig.1(b) $K_\epsilon=0.04$,$T=2$ and 
$g\approx 9.5$. As the resonance is approached the amplitude of the 
oscillations is proportional to the
square of their wavelength, \ie 
$E(t=NT) \propto {K^4} \frac{1}{(\delta)^2} \sin^2 N\delta$
where $2\delta$ is the distance from the resonance peak. 
{\bf(c)} Corresponding number of non-condensate atoms
from Eq.\ref{bogo}.}
\label{Fig4}
\end{figure}

\section{IV: Bogoliubov resonances for $T=2\pi$}

The kick period $T=2\pi$, in a non-interacting system of cold atoms 
(\ie $g=0$) corresponds to a so-called ``quantum anti-resonance''
where the cold atom cloud exhibits periodic (period-2) oscillations.
Hence the isolated Bogoliubov resonance regime to  higher 
$K$ than would be expected for generic $T$.
The effect of a non-zero $g$ for  $T=2\pi$ was investigated
in \cite{Zhang}. An instability  border  occurring at a critical 
value of nonlinearity, \eg for $g \simeq 2$ at $K=0.8$,
was identified where the growth on non-condensate particles with time 
became exponential. \\

In Fig.\ref{Fig4}(a) we investigate the behavior near critical $g$, 
for $K=0.8$. We see that if a wider range of $g$ is considered, 
the stability border is also a resonance: the condensate rapidly 
recovers stability after the instability border is passed. 
The condensate is exponentially unstable for $g \simeq 2 \to 2.6$,
but is quite stable for both $g=1.5$ and $g=3$, as shown. Fig.\ref{Fig4}(b)
shows oscillations in the condensate energy, as a function of time; 
a smoothed plot is also shown. For $g=1.5$ and $g=2.8$ (off-resonance) 
the smoothed plots are flat; for $g=2.2$ and $g=2.5$ (near-resonant), 
slow deep oscillations are apparent.   

The behavior is analogous to that of generic $T$; however,
the analysis of the condensate resonances for $T=2\pi$ is less 
straightforward: the strongest resonances, even for low $K \lesssim 2$, 
do not in fact occur for $ \omega_l T \approx 2\pi M$, where $M=1,2,3...$.

A significant difference between generic $T$ and $T=2\pi$ is that, for 
the generic case, if we write
\begin{equation}
\omega_l T \approx 2\pi M_l + 2\delta(l)
\end{equation}
we see that for arbitrary generic $T$, the distance from the
nearest resonance, for the different modes, depends on $l$.
In contrast, for $T=2\pi$, for large $l$ (\ie  $l \gtrsim 3$)
we find $\omega_l T \approx (l^2 + \frac{g}{\pi}) \pi $; in 
other words, the de-phasing from the nearest resonance (and hence
the period of the mode oscillations) is
similar (either $2\delta \approx \frac{g}{\pi}$ or 
$2\delta \approx 1- \frac{g}{\pi}$)
for for all modes. So all mode oscillations for high $l$ 
are approximately in phase with each other. 

For $K=0.8$, only low modes $l=1,2$ are significantly populated.
These low modes ($l=1$ and $l=2$) are only in phase with each other 
at certain precise values of $g,T$. For these parameters, the 
model of Eq.\ref{map} predicts large
resonances whenever the condition $(\omega_1+ \omega_2) T \approx 2\pi M$
is satisfied. In particular, for the resonance near $g \approx 2$,
we find that for the $l=1$ mode, $\omega_1 T \approx (1- 2\delta) 2\pi$ while
for the $l=2$ mode $\omega_2 T \approx (2+ 2\delta) 2\pi$, 
with $2\delta \approx .25$.

These results suggest that ``two-mode resonances'', \ie synchronized 
oscillations of pairs of the lowest excited modes are the 
dominant mechanism for $T=2\pi$ (NB this could be viewed as a ``three-mode''
resonance, if we include the lowest, initial mode, but $\omega_0=0$ for
our system).
They account for the shifting position of the critical instability
border found by \cite{Zhang} in the $T=2\pi$ case. For example, 
for slightly higher kick strengths, such as $K\simeq 2$, a
resonance appears for $g \approx 1.65$ corresponding to 
$(\omega_2+ \omega_3) T \approx 2\pi M$,
which accounts for the displacement of the instability border to lower 
values of $g$. Note that the resonance positions in the full numerics 
are $K$-dependent,
whereas in the perturbative model of Eq.\ref{map} this dependence is neglected; 
the model is only valid for very small $K$.

\begin{figure}
\includegraphics*[width=3.in]{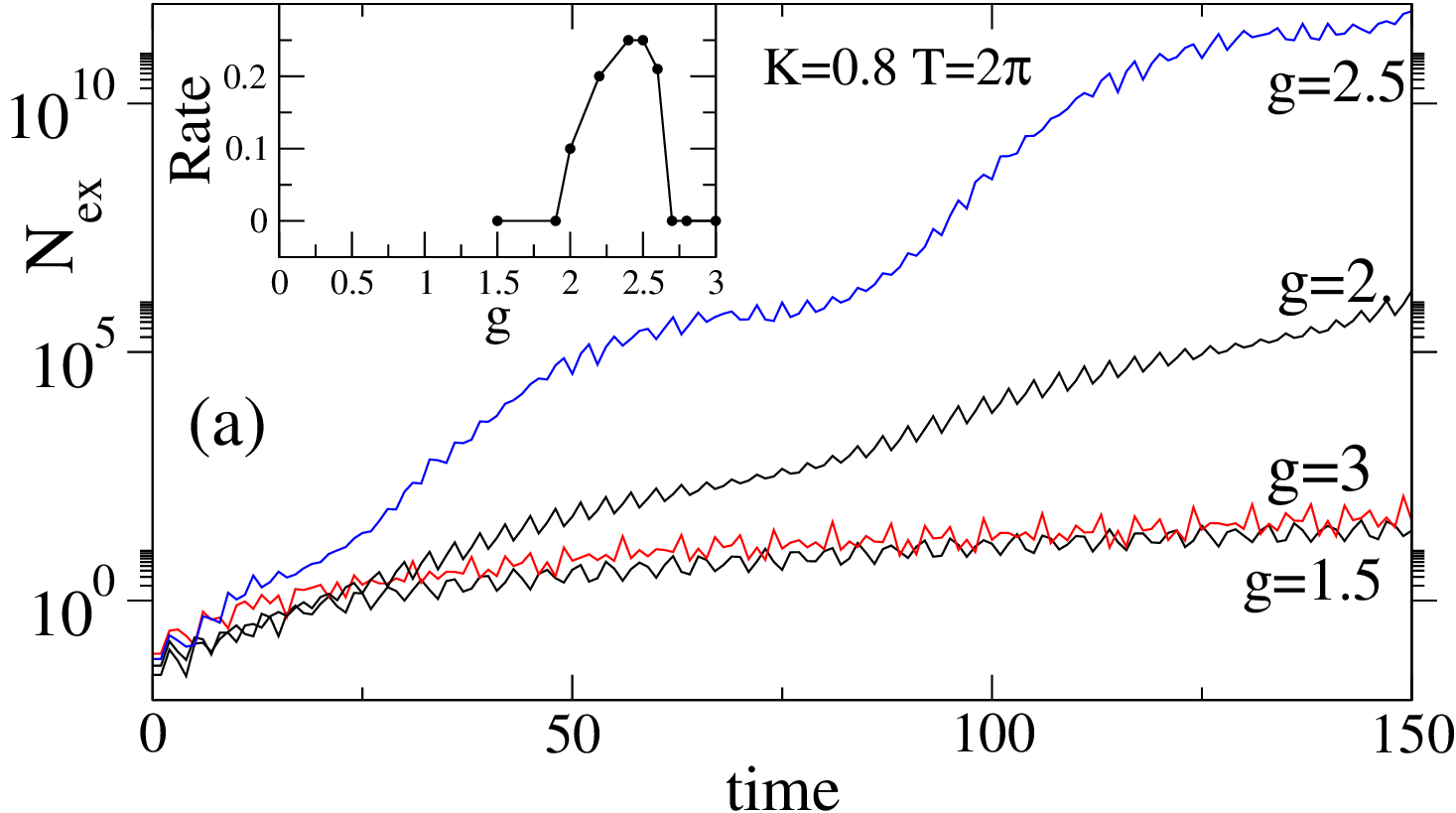}
\vspace*{10 mm}
\includegraphics*[width=3.in]{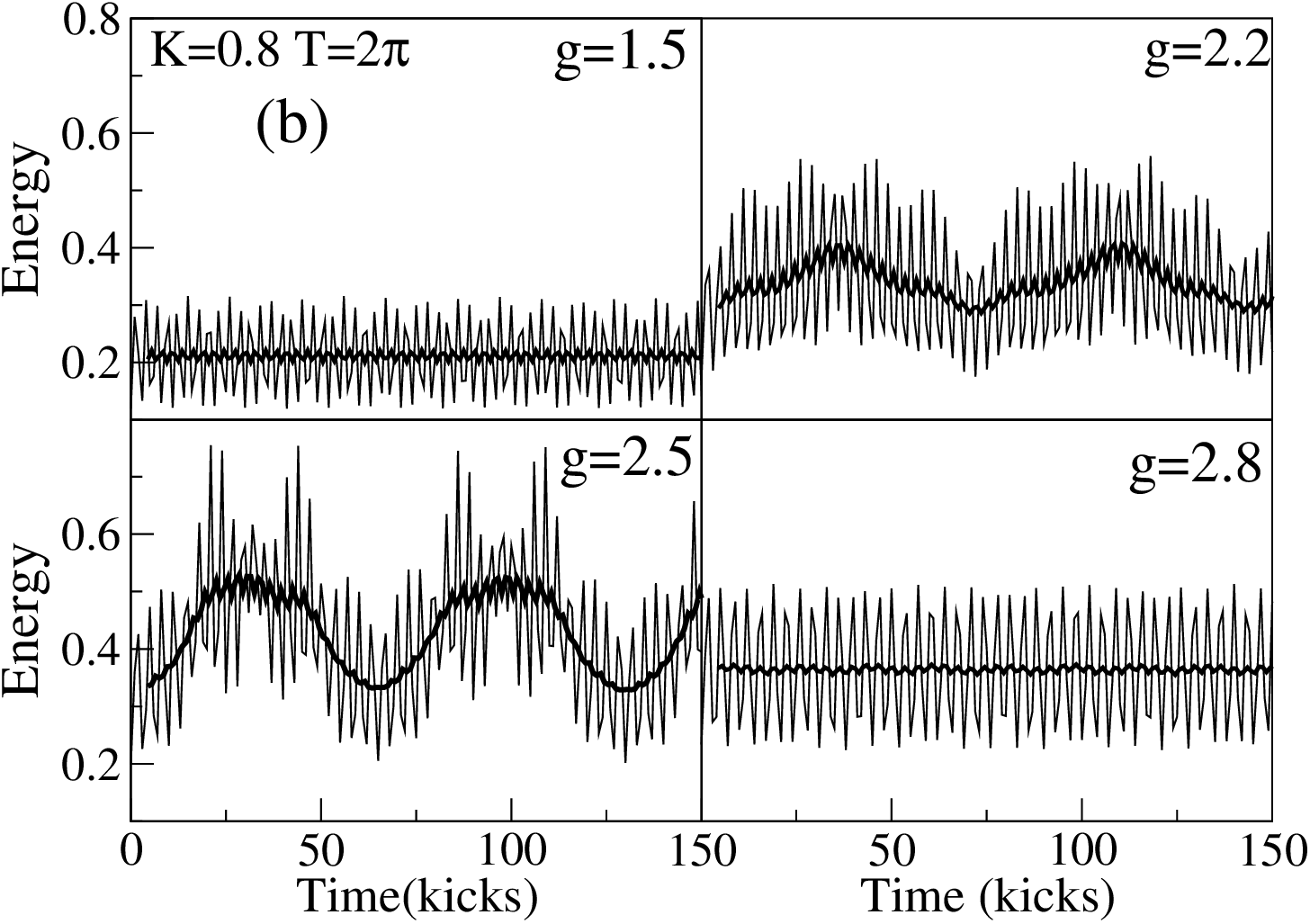}
\caption{(a) Non-condensate particles for kicking period 
$T=2\pi$, $K=0.8$, $g \simeq 2$. The inset shows 
the rate of exponential growth of non-condensate atoms; zero denotes 
polynomial growth or less. The graph shows this instability border 
is a resonance: the condensate is unstable for $g=2-2.5$ but is
stable for $g=1.5$ and $g=3.0$. 
(b) Energy oscillations as a function of
time; smoothed plots are also shown. 
Before and after the resonance ($g=1.5$ and $g=2.8$) the smoothed
plots are flat. Near-resonance, ($g=2.2$ and $g=2.5$) the energy 
shows the characteristic slow, deep resonant oscillations.}
\label{Fig5}
\end{figure}

\section{V: Conclusion}
In conclusion, 
we have shown that exponential instability in kicked BECs is related
to parametric resonances, ie driving of low-lying collective modes
at their natural frequencies, rather than to
chaos in the underlying mean-field dynamics \cite{gard2}.

 The signature of this process
is in the onset of slow, large amplitude periodic oscillations in the
condensate energy as well as the number of non-condensate atoms 
calculated from the time-dependent Bogoliubov formalism, as a resonance 
is approached. The resonances proliferate and overlap
for large kick-strengths $K$, leading to instability over wider ranges of
$K$ and $g$. 
The time-dependent Bogoliubov approximation used here and in all
other previous studies is only valid in regimes where the condensate
depletion is negligible; for realistic condensates analysis of the
dynamics in the narrow (for weak driving)  windows of parametric instability,
would require other approaches beyond Bogoliubov.
 However, away from these windows, the kicked condensate remains
stable and relatively unperturbed, even after prolongued kicking.

\bigskip

JR acknowledges funding from an EPSRC-DHPA scholarship.
The authors would like to thank Chuanwei Zhang  for valuable advice.
This research was supported by the EPSRC.


\begin{thebibliography}{99}
\bibitem{Stringari}
{L.P.~Pitaevskii and S.~Stringari, {\it Bose-Einstein Condensation}
(Oxford University Press, Oxford, 2003);F.Dalfovo, S.Giorgini,
L.P.~Pitaevskii, S Stringari, Rev.Mod.Phys.{\bf 71} 463 (1999)}
\bibitem{Shep} 
{D.L.~Shepelyansky, Phys. Rev. Lett. {\bf 70}, 1787 (1993).}
\bibitem{Gardiner} 
{S.A.~Gardiner, D.~Jaksch, R.~Dum, J.I.~Cirac, and P.~Zoller,
Phys. Rev. A {\bf 62}, 023612 (2000);
R.~Artuso and L.~Rebuzzini, Phys. Rev. E {\bf 68}, 036221 (2005).}
\bibitem{Garreau}
{Q.~Thommen, J.C.~Garreau, and V.~Zehnle, Phys. Rev. Lett. {\bf 91}, 
210405 (2003).}
\bibitem{Duffy}
{G.J.~Duffy, A.S.~Mellish, K.J.~Challis, and A.C.~Wilson,
Phys. Rev. A {\bf 70}, 041602(R) (2004).}
\bibitem{Zhang}
{C.~Zhang, J.~Liu, M.G.~Raizen, and Q.~Niu,
Phys. Rev. Lett. {\bf 92}, 054101 (2004).}
\bibitem{Zhang2}{J.~Liu, C.~Zhang M.G.~Raizen, and Q.~Niu, Phys. Rev. A.
{\bf 73} 013601, (2006).}
\bibitem{Wimberger}
{S.~Wimberger, R.~Mannella, O.~Morsch, and
E.~Arimondo, Phys. Rev. Lett. {\bf 94}, 130404 (2005).}
\bibitem{Adams}
{A.D.~Martin, C.S.~Adams, and S.A.~Gardiner, Phys. Rev. Lett.
{\bf 98}, 020402 (2007).}
\bibitem{Raizen} 
{M.G.~Raizen, Adv. At. Mol. Opt. Phys. {\bf 41}, 43 (1999).}
\bibitem{Darcy}
{M.K.~Oberthaler, R.M.~Godun, M.B.~d'Arcy, G.S.~Summy, and K.~Burnett,
Phys. Rev. Lett. {\bf 83}, 4447 (1999).}
\bibitem{Fish} {S.~Fishman, I.~Guarneri, and L.~Rebuzzini, Phys. Rev. Lett
{\bf 89}, 0084101 (2002); L. Rebuzzini, S.Wimberger and R.Artuso,
 Phys.Rev.A {\bf 71}, 036220 (2005).}
\bibitem{Rebuzzini} L. Rebuzzini, R.Artuso, S.Fishman, I.Guarneri, Phys. Rev. A
 76, 031603 (2007).
\bibitem{Sadgrove}
{C.~Ryu, M.F.~Andersen, A.~Vaziri, M.B.~d'Arcy, J.M.~Grossman, 
K.~Helmerson, and W.D.~Phillips, Phys. Rev. Lett. {\bf 96}, 160403 (2006); 
M.~Sadgrove, M.~Horikoshi, T.~Sekimura, and K.~Nakagawa, 
Phys. Rev. Lett. {\bf 99}, 043002 (2007); G. Behinaenin,
V.Ramareddy,P.Ahmadi and G.S.Summy,  Phys. Rev.
Lett.{\bf 97}, 244101 (2006); I.Dana, V. Ramareddy, I. Talukdar and G. S. Summy,
 Phys. Rev. Lett. 100, 024103 (2008); J. F. Kanem, S. Maneshi, M. Partlow M. Spanner and A. M. Steinberg
 Phys. Rev. Lett. 98, 083004 (2007).}
\bibitem{Wu} B.Wu and Q. Niu New J. Phys. {\bf 5}, 104 (2003) 
\bibitem{Dalfovo} M.Kramer, C.Tozzo, F.Dalfovo, Phys. Rev. A {\bf 71}, 061602(R) (2005);
 Phys. Rev. A {\bf 72}, 023613 (2005).
\bibitem{Poletti} 
{D.~Poletti, G.~Benenti, G.~Casati, and B.~Li,
Phys. Rev. A {\bf 76}, 023421 (2007).}
\bibitem{PRTHEO}J.J. Garcia-Ripoll, V.M.Perez-Garcia and P Torres,
Phys. Rev. Lett. {\bf 83}, 1715 (1999);Yu Kagan and L A Manakova, cond-mat/0609159 (2006). 
\bibitem{PREXPT} N.Gemelke et al, Phys. Rev. Lett. {\bf 95}, 170404 (2005);
G.Campbell et al, Phys. Rev. Lett. {\bf 96}, 020406 (2005); P.Engels, C.Atherton and 
M.A.Hoefer Phys. Rev. Lett. {\bf 99} 095301 (2007). 
\bibitem{Jones} 
{P.H.~Jones, M.~Stocklin, G.~Hur, and T.S.~Monteiro,
Phys. Rev. Lett. {\bf 93}, 223002 (2004);
 C.E. Creffield, G. Hur, T.S. Monteiro, Phys. Rev. Lett. {\bf 96}, 024103 (2006);
J. Wang, T S Monteiro, S.Fishman, J.Keating, R.Schubert Phys. Rev. Lett. {\bf 99}, 234101 (2007).}
\bibitem{rescalg} 
{This means that we work with an effective value of the nonlinearity $g$.}
\bibitem{castin}
{Y.~Castin and R.~Dum, Phys. Rev. Lett. {\bf 79}, 3553 (1997);
Phys. Rev. A {\bf 57}, 3008 (1998).}
\bibitem{gard} S.A.Gardiner and S.A.Morgan, {\bf 75}, 043621 (2007).
\bibitem{gard2}While we can not draw any conclusions on the Kicked Harmonic Oscillator,
as we do not study it, 
we note that eg Figs 18 and 19 in the study in [3] show deep slow oscillations 
 suggestive of an approach to a Bogoliubov resonance (not necessarily
leading to exponential behavior in those examples). 

\end{thebibliography}
\end{document}